\documentclass[%
 aip,
 amsmath,amssymb,
 reprint,%
]{revtex4-1}

\usepackage{graphicx}% Include figure files
\usepackage{dcolumn}% Align table columns on decimal point
\usepackage{bm}% bold math
\usepackage[utf8]{inputenc}
\usepackage[T1]{fontenc}
\usepackage{mathptmx}

\usepackage[usenames,dvipsnames]{color}

\begin{document}

\preprint{AIP/123-QED}

\title{The Blind Implosion-Maker -\\ Automated Inertial Confinement Fusion experiment design}
% Force line breaks with \\

\author{P.W.Hatfield}
% \altaffiliation[Also at ]{Clarendon Laboratory, University of Oxford, Parks Road, Oxford OX1 3PU, UK}%Lines break automatically or can be forced with \\
 \email{peter.hatfield@physics.ox.ac.uk}
\affiliation{%
Clarendon Laboratory, University of Oxford, Parks Road, Oxford OX1 3PU, UK
}%


\author{S.J.Rose} \altaffiliation[Also at ]{Clarendon Laboratory, University of Oxford, Parks Road, Oxford OX1 3PU, UK}%Lines break automatically or can be forced with \\
\affiliation{
Blackett Laboratory, Imperial College, London SW7 2AZ, UK%\\
}%

\author{R.H.H.Scott}
\affiliation{
Central Laser Facility, STFC Rutherford Appleton Laboratory, Harwell Oxford, Didcot OX11 0QX, UK%\\
}

\date{\today}% It is always \today, today,

\begin{abstract}
The design of inertial confinement fusion experiments, alongside improving the development of energy density physics theory and experimental methods, is one of the key challenges in the quest for nuclear fusion as a viable energy source\cite{Hurricane2016}. Recent challenges in achieving a high-yield implosion at the National Ignition Facility (NIF) have led to new interest in considering a much wider design parameter space than normally studied\cite{Peterson2017}. Here we report an algorithmic approach that can produce reasonable ICF designs with minimal assumptions. In particular we use the genetic algorithm metaheuristic, in which `populations' of implosions are simulated, the design of capsule is described by a `genome', natural selection removes poor designs, high quality designs are `mated' with each other based on their yield, and designs undergo `mutations' to introduce new ideas. We show that it takes $\sim 5\times10^{4}$ simulations for the algorithm to find an original NIF design. We also link this method to other parts of the design process and look towards a completely automated ICF experiment design process - changing ICF from an experiment design problem to an algorithm design problem.
\end{abstract}

\maketitle

\section{\label{sec:level1} Introduction}

Achieving controlled nuclear fusion burn in the laboratory is a key goal in the pathway to nuclear  fusion as an industrial power source. One of the main potential pathways to this goal is inertial confinement fusion (ICF), in which deuterium-tritium fuel is compressed to extremely high temperatures and pressures very quickly. The world's leading ICF facility is NIF, in California, USA, at Lawrence Livermore National Laboratory, although ICF is a worldwide endeavour with facilities around the world contributing towards this goal.

When considering different designs, normally the intention is to maximise the neutron yield. In this sense ICF can be viewed as an optimisation problem\cite{Giunta2002}: we wish to maximise yield within the constraints of the experimental setup (e.g. within what experiments could feasibly be fielded on NIF). The design space of ICF experiments is very large, making exploration of the design space non-trivial. Designs have historically been found by human imagination combined with understanding of physical principles. However the inability of conventional designs to reach ignition, combined with the explosion of applications of algorithmic and machine learning techniques in the physics community at large, has led to an increase in the use of computer algorithms and machine learning approaches to identify new designs\cite{Baltz2017,Peterson2017,Humbird2018}.

In this work we consider a much larger parameter space than hitherto, and seek to design an ICF experiment completely `from scratch' (as opposed to within some pre-defined smaller parameter space). In particular this work is motivated by the `The Surprising Creativity of Digital Evolution'\cite{Lehman2018} - which summarises a variety of unexpected results from a range of fields, where algorithmic approaches produced completely unanticipated designs. For ICF at NIF it would be of great interest to find if there are any `unexpected' design features that are yet to be found that could help reach ignition; conversely if such features don't exist, we would like to increase our confidence that no such alternative designs do exist.

\section{\label{sec:level2} Genetic Algorithms}

We consider the performance of the class of metaheuristics known as \textit{genetic algorithms}\cite{Holland1992} (GAs), which are motivated by Darwinian evolution, and have already been applied to some problems in fusion and laser physics\cite{Miner2001,He2015}. One particularly relevant experimental campaign to this analysis is an effort to optimise x-ray production from an argon cluster jet at the Gemini laser facility\cite{Streeter2018}. They found that they were able to double the x-ray production from their experiments with a genetic algorithm - with comparatively little knowledge of the underlying physics.

\begin{figure*}[h!]
\includegraphics[scale=0.6]{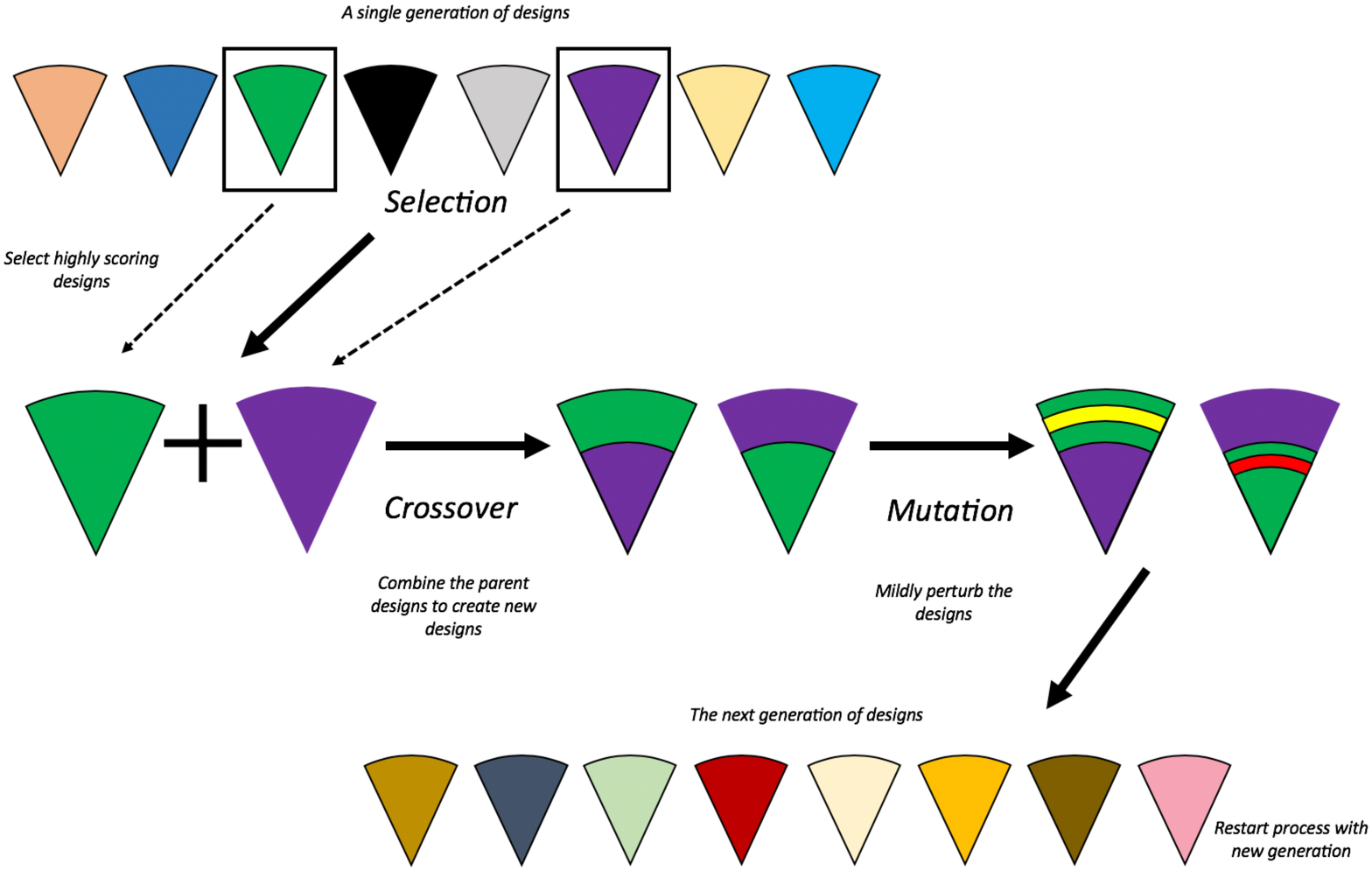}
\caption{\textbf{The Genetic Algorithm} Schematic of the key components of the genetic algorithm as applied to the capsule designs. A similar process is applied to the drive designs.}
\label{fig:cartoon}
\end{figure*}

In genetic algorithms, the design in question is characterised by a `genome'.  A `population' of designs is simulated, the highly performing ones are given higher chances of `mating' with each other (crossover), receive `mutations' and produce a new generation. We use a crossover method that respects the capsule (and drive) structure, see figure \ref{fig:cartoon}. This process is then repeated indefinitely, producing better and better designs. Genetic algorithms are typically most appropriate when little is known about the parameter space - different algorithms may be better if more is known about the problem. The `No Free Lunch' theorem\cite{Wolpert1997} suggests that a perfect `black-box' optimiser likely cannot exist, so in general an appropriate algorithm for the situation must be chosen, and performance between algorithms may need to be tested empirically.

Our implosion simulations are performed using the {\sc Hyades}\cite{Larsen1994} radiation-hydrodynamics simulation code, which is well benchmarked and used widely for the simulation of inertial fusion and high energy density physics applications\cite{Keilty2000,Leibrandt2005,Sequoia2006, Scott2012}. {\sc Hyades} models hydrodynamics within a Lagrangian framework. Electron and ion thermal energy transport is described by a flux-limited Spitzer-H{\"a}rm thermal conductivity model. Equations-of-state either use the Los Alamos SESAME tables\cite{Hu2011} or QEOS\cite{More1988}. Ionization levels come from a hydrogenic average-atom model or self-consistently from QEOS. Radiation transport uses the multi-group diffusion approximation; here we use 60 groups. A 1D spherically symmetric geometry is employed.

Neither laser-plasma interactions nor hohlraum physics are modelled; instead we use an incoming x-ray drive imposed at the outside of the grid. The capsules are modelled within a 5mm helium container. Simulations start in cryogenic conditions at $1.551\times10^{-3}$ev$=18$K. The {\sc Hyades} runs were performed on SCARF at the Central Laser Facility at Rutherford Appleton Laboratory using 600 CPUs.

The design space is chosen to roughly correspond to what is achievable with a gold hohlraum on NIF. We consider capsule designs made of 5 layers. Each layer can be DT gas, DT ice, or plastic (CH). If the layer is DT gas, its density may be between 10mg/cc and 200mg/cc. Note this lower bound is substantially higher than the DT gas density in the NIF point design, so we would not expect it to be possible to achieve the higher yields possible within other design spaces. This constraint is motivated by a desire to avoid implosions with extremely high convergence ratios; partially because 1D simulations are likely to be more realistic for low convergence ratio implosions, and partially because there is some suggestion that the reliance on very high convergence ratios is contributing to the difficulty in achieving ignition. DT ice and CH densities are fixed at 255mg/cc and 1,044mg/cc respectively. The thickness of each layer may be between 0.01mm and 1.1mm. The total radius of the capsule is constrained to be less than 1.1mm. For the x-ray drive a pure Planckian spectrum (no M-band) is used, and its time dependence is modelled by combining 15 Gaussians, each of which is parametrised by three numbers (peak temperature, time of peak and width). The drive is constrained to never go above 300ev, and to not deliver more than $300$kJ to a 1mm radius sphere, and its cooling rate is constrained to be below 50ev/ns, all of which are considered realistic constraints for capsule implosion in a hohlraum at NIF. Rigorous modelling of hohlraum physics to determine precisely what drives are possible is beyond the scope of this text, but these rough constraints are consistent with other experiments fielded at the facility\cite{Hurricane2016,Hurricane2014,Ping2019}. This is a very large parameter space: of order 60 discrete or continuous parameters.

\begin{figure*}
\includegraphics[scale=0.8]{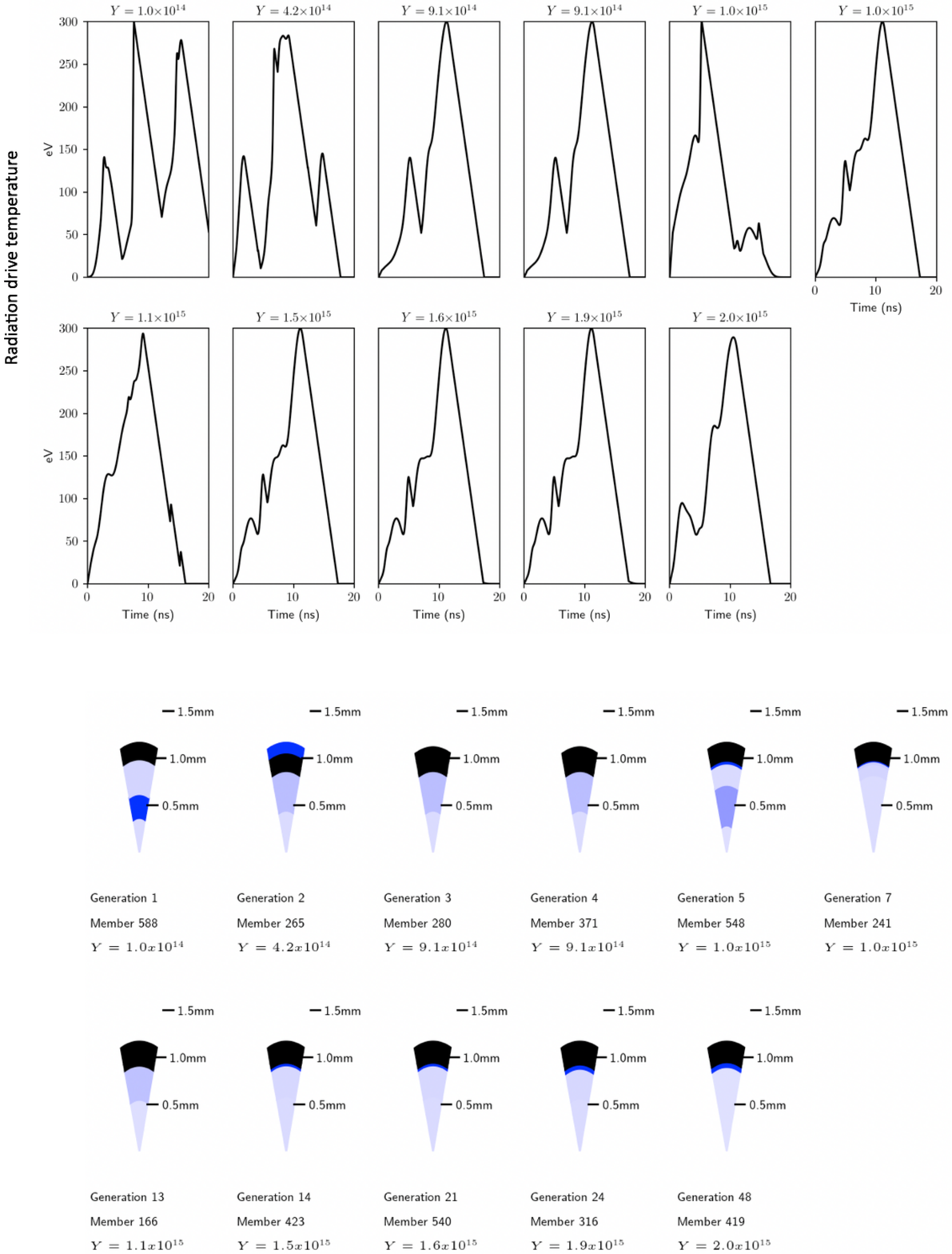}
\caption{\textbf{Survival of the fittest - designs over time.} Top panel shows capsule designs and the bottom panel shows the corresponding drive designs. Designs only shown when a new generation improves over the previous generation. Pale blue represents DT gas (with darker corresponding to higher densities), royal blue to DT ice, and black to CH.}
\label{fig:designs_over_time}
\end{figure*}

\begin{figure}
\includegraphics[scale=0.8]{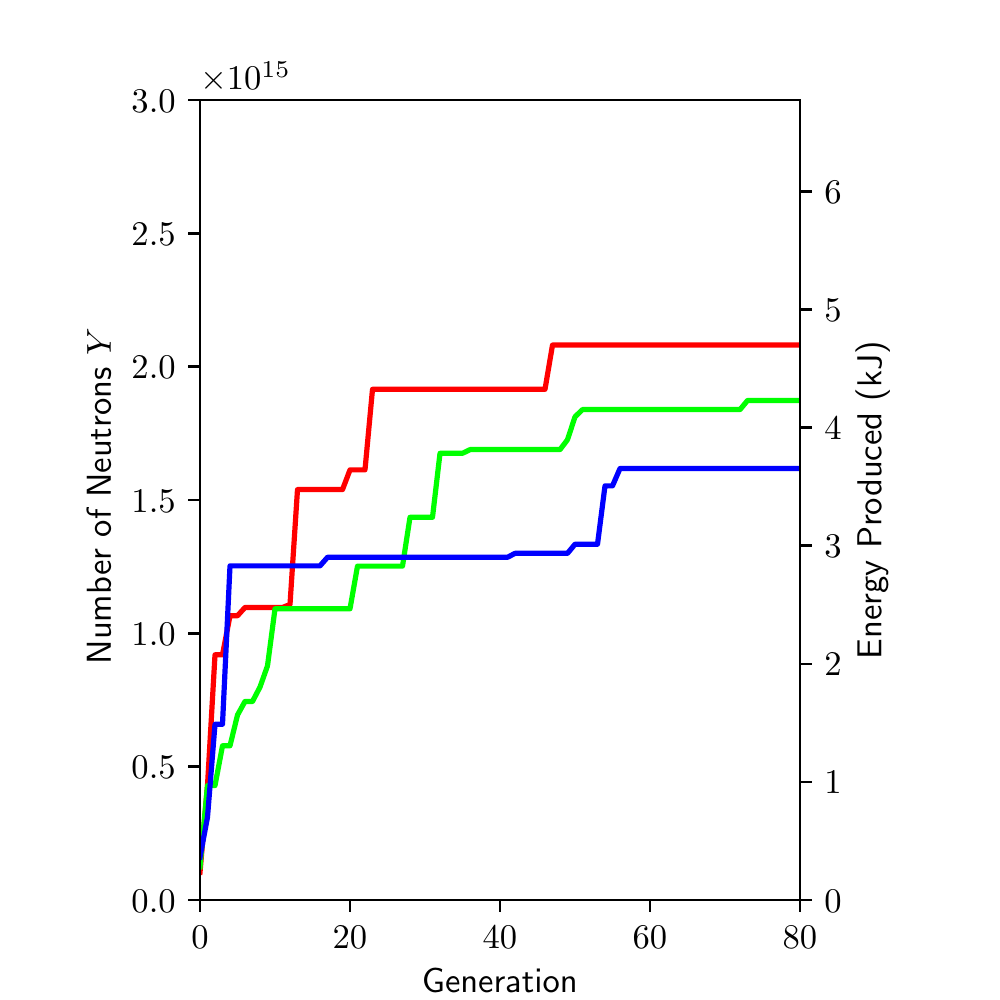}
\caption{\textbf{Yield improvement with number of generations.} The highest yield achieved as a function of population generation.}
\label{fig:yield_with_gen}
\end{figure}

\begin{figure}
\includegraphics[scale=0.7]{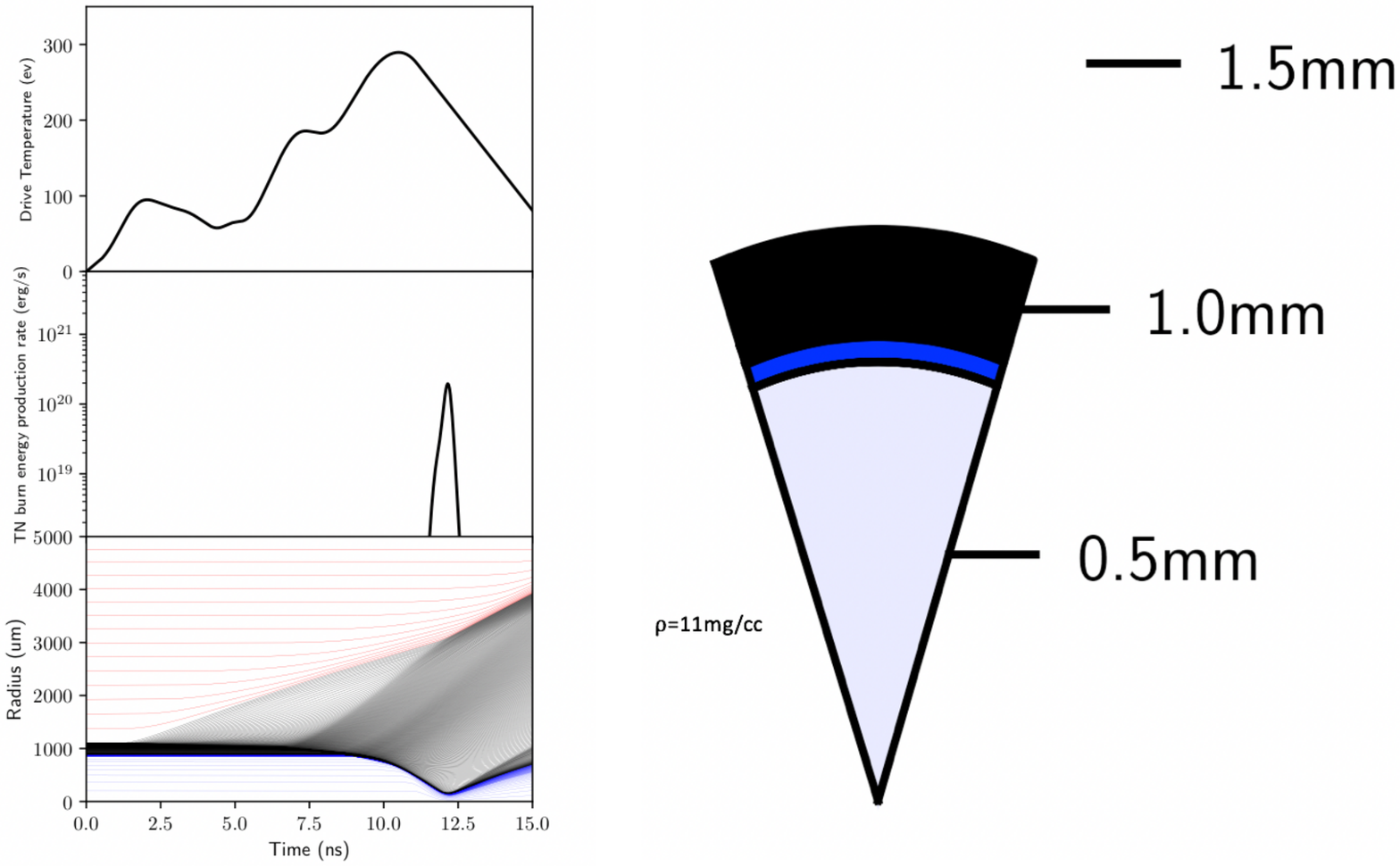}
\caption{\textbf{Structure of the final implosion.} Top panel of the left half of the figure shows the radiation drive for the final design as a function of time, middle panel shows thermonuclear burn energy rate as a function of time, bottom panel shows a radius-time plot of the final implosion. Pale blue represents DT gas (with darker corresponding to higher densities), royal blue to DT ice, black to CH and red to He. The right half of the figure shows the final capsule design.}
\label{fig:time_radius}
\end{figure}

A population size of 600 was run for 80 generations: 48,000 {\sc Hyades} simulations in total, taking $\sim$27 days in total to run, and finding a recognisable design. The evolution in drive and capsule structure leading up to the final design is shown in figure \ref{fig:designs_over_time}, and corresponding increase in yield in figure \ref{fig:yield_with_gen}. The time-radius profile of the final implosion is shown in figure \ref{fig:time_radius}. The algorithm was run three times (each time with different seeds), and the best performing final design had a total neutron yield of $\sim 2.1 \times 10^{15}$ neutrons ($\sim$5 kJ). This  yield is achieved with a modest convergence ratio. We note that in future implementations the design need not be optimised over yield. For example, instead of being treated as a constraint, the total energy in the drive could be minimised, with the constraint of $Y>10^{18}$ (say) imposed.

\begin{figure*}
\includegraphics[scale=0.6]{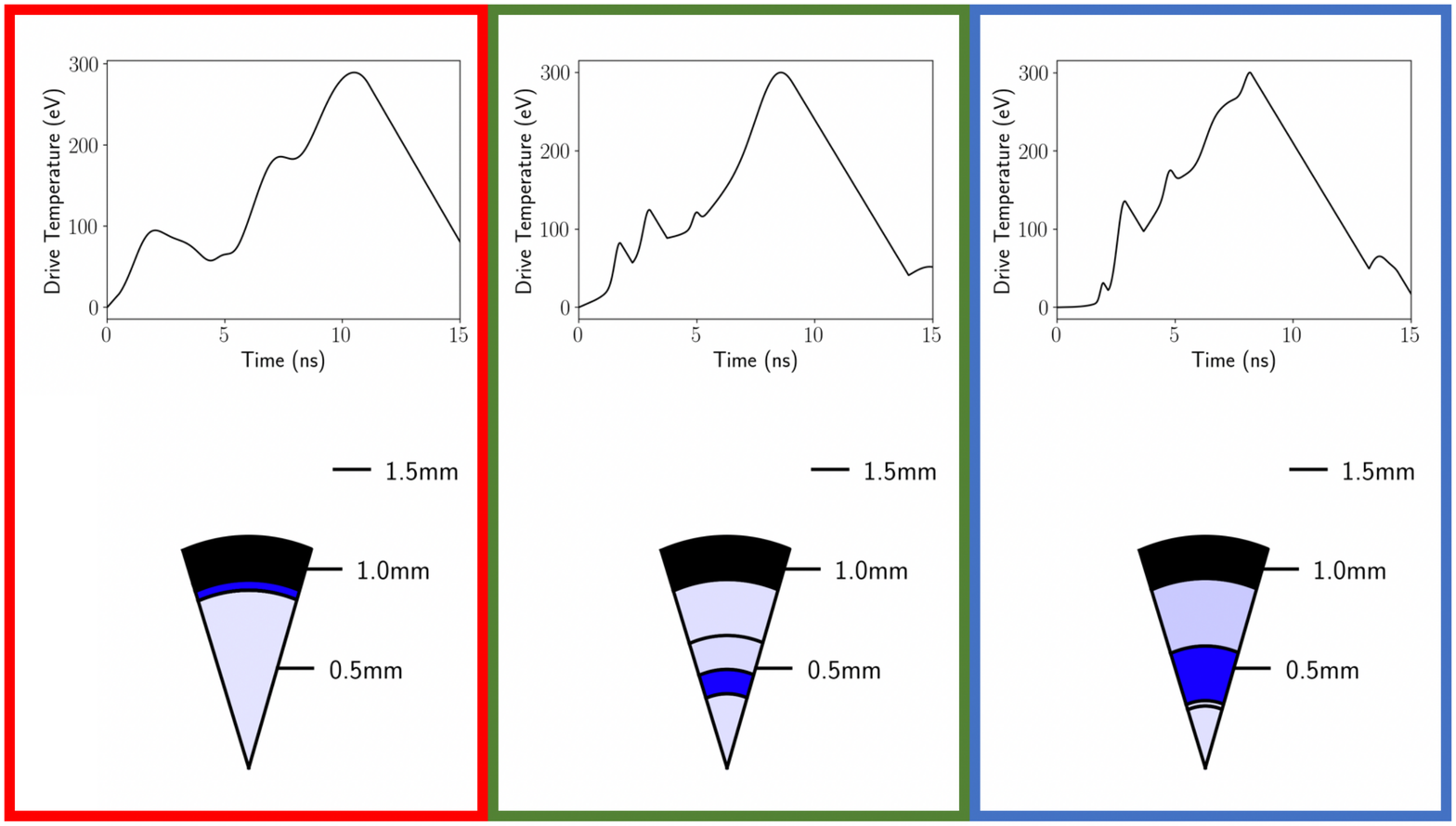}
\caption{\textbf{Different Designs} The final capsule and drive designs of each of the runs of the algorithms, box colours match to figure \ref{fig:yield_with_gen}.}
\label{fig:multiple_designs}
\end{figure*}

In terms of algorithmic convergence, each run did not converge to exactly the same design, see figure \ref{fig:multiple_designs}. All three achieve a factor of ten improvement very rapidly in the first $\sim5-10$ generations, and then get a further 50\%-100\% improvement in the subsequent 70 generations much more slowly. Smaller mutations likely would make the initial rapid increase in yield with generation slower, and would increase the risk of being stuck in a local maximum, but would let the fine tuning of the design proceed more quickly at later stages. This shows that the GA, as implemented here, does not achieve the global maximum every time (which is likely very hard in such a large parameter space), but does consistently find plausible designs in the region of what is likely to be the global optimal. All three designs have similar ablators, and a similar peak to the drive with some pre-shocks, and a mix of DT gas and ice in the interior of the capsule (in fact it is interesting that it looks like it may be possible to get comparable yields when the ice is not on the inner surface of the ablator). The final yields are within 25\% of each other, reasonably similar considering the $\sim 20$-fold improvement that is found from the first generation. It is also in general quite hard to judge when the algorithm has converged; for the `blue' run it might have been tempting to stop the algorithm at around 40 generations.  More fundamentally  however, finding the absolute global maximum is not the intention; GAs are known to be very slow in the final stages of convergence, but critically we would not expect the global maximum in this simplified 1D setup to be the maximum for real experiments. The intention is to find plausible outlines of designs in an acceptable amount of time, that can then be further refined; see section \ref{sec:level3}. Large amounts of computational resources could be wasted trying to find the absolute maximum in 1D, when it is likely that such a design would not be the experimental optimal.

\section{\label{sec:level3} Automating Design}

\begin{figure}
\includegraphics[scale=0.4]{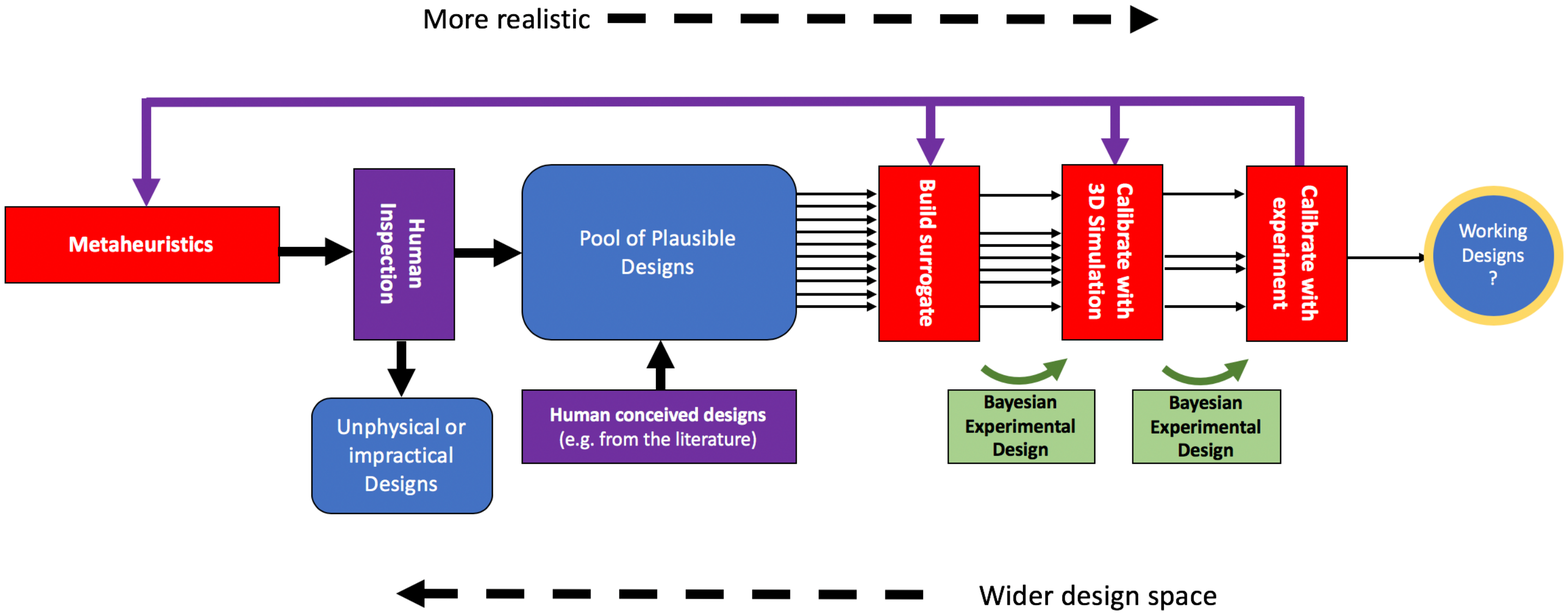}
\caption{\textbf{Schematic of `The Data Scientists' Algorithm'.} Blue boxes show pools of designs, purple sections and lines show processes requiring human input, red boxes are algorithmic processes, green lines and boxes show decision making processes, and black lines show the flow of designs}
\label{fig:flow_chart}
\end{figure}

Although the design found here is worthy of greater analysis, the key result is that finding plausible ICF designs can be automated. Studies similar to the analysis presented here will likely (for the foreseeable future) only be able to do the large number of simulations presented here in 1D and with simplified physics. For this reason we view metaheuristics as one stage in the `Data Scientists' Algorithm' approach to ICF design. Figure \ref{fig:flow_chart} shows a schematic for this alternative approach where metaheuristics are used to generate basic designs\cite{Kittara2007,Hector2017}. Human inspection removes designs that are clearly unphysical or impractical and reduces the dimensionality of the designs to make the subsequent steps tractable. These designs then form a `pool' of plausible designs alongside those that are produced by humans. Each of these designs is then studied in greater depth using machine learning regression based surrogates\cite{Peterson2017} and designs that start to fail when more realistic physics is included can be removed (e.g. perhaps a design had high yield in 1D but completely fails in 2D). These surrogates can then be calibrated with more computationally expensive state-of-the-art 3D simulations (where it would be computationally unfeasible to do more than a few full calculations; in addition the increase in fidelity need not be 1D to 3D, it could also be single to multi-group diffusion etc.), and then calibrated again with real NIF shots\cite{Humbird2018,Gaffney2018,Gopalaswamy2019}. These experimentally calibrated surrogates give data-based improved predictions over theory without learning why theory and experiment disagree in a physical way. However if new physical insight is learnt from the experiments then it can be fed back into the simulations used in the process. The experiments used to calibrate the surrogate should be chosen using Bayesian Experimental Design/Bayesian Optimisation\cite{Marco2017,Dieb2018}.

\begin{figure}
\includegraphics[scale=0.5]{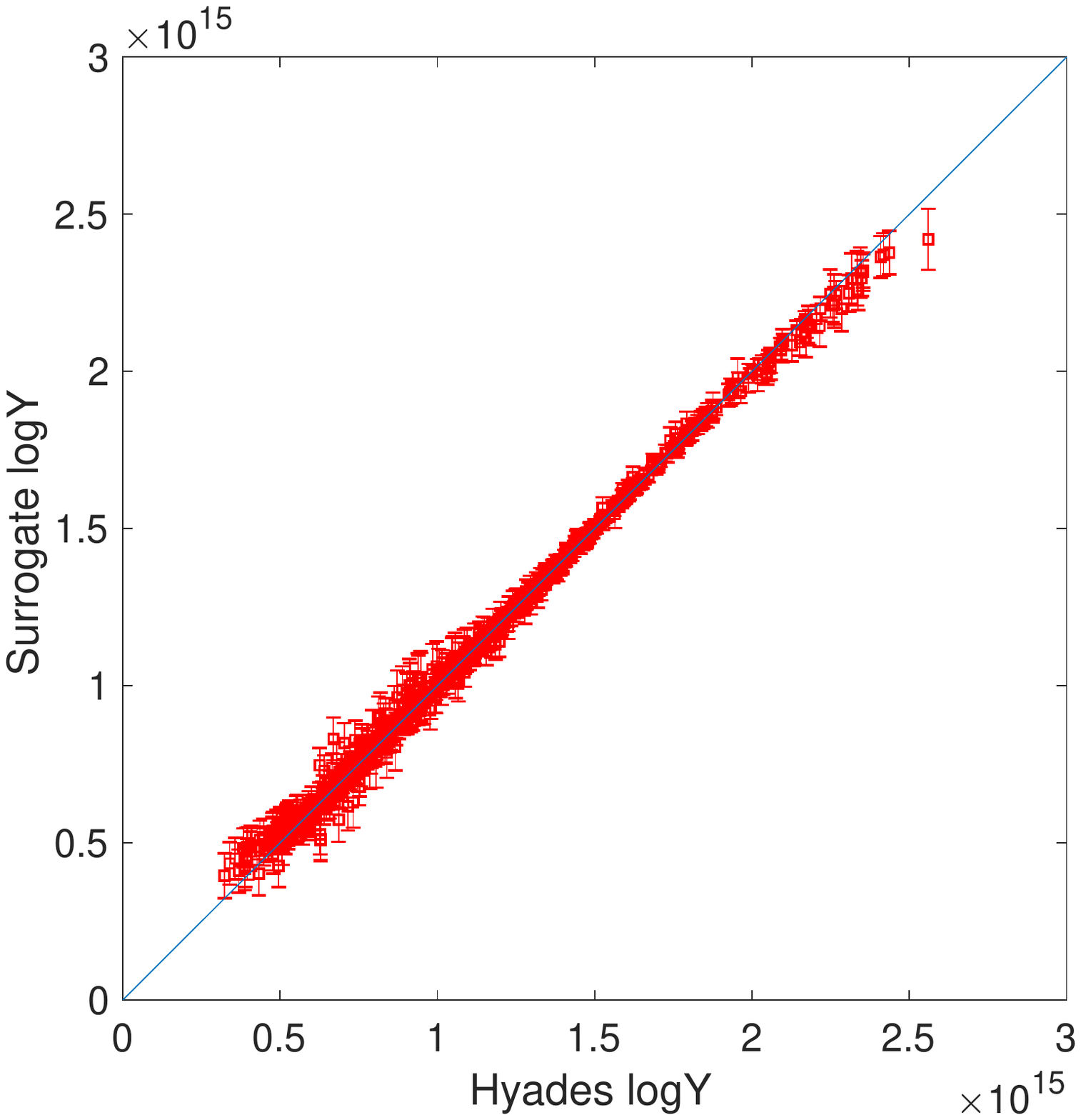}
\caption{\textbf{Surrogate predictions} The graph shows the performance of our machine learning surrogate, the x-axis shows the {\sc Hyades} yield and the y-axis shows what the machine learning surrogate predicted (with uncertainties) for that point in design space. Light blue diagonal line shows a one-one relationship.}
\label{fig:ML_example}
\end{figure}

We give a brief example of how this proposed workflow might work. Our genetic algorithm metaheuristic earlier gave three designs. Human inspection would note that the `green' and `blue' designs are relatively similar, and the 'red' design slightly different. A human designer might in this case  judge that the red design is more worthy of further testing, both as it had a higher yield, and that it had fewer potential engineering difficulties than the green/blue design. Once the red design was selected, the human designer might then judge that varying the gas fill density and the ablator thickness were the most important parameters and start to apply a surrogate building methodology\cite{Peterson2017}. We demonstrate this by doing 1000 simulations over this 2D parameter space, letting gas fill density vary between 10mg/cc and 30mg/cc, and ablator thickness vary between 0.1mm and 0.3mm (keeping total radius and DT ice thickness fixed but also letting the DT gas radius vary in tandem). We then applied the machine learning code GPz\cite{Almosallam2016a,Almosallam2016b,Hatfield2019} to this data (100 data points in the training sample, 100 in the validation sample, and 800 in the testing sample); results in figure \ref{fig:ML_example} show that this modest training set is sufficient for this small parameter space. With the surrogate one can a) see that a slightly higher yield is indeed possible (an improvement from $\sim 2.1 \times 10^{15}$ to $\sim 2.5 \times 10^{15}$ neutrons) and b) now with the ability to quickly predict yield in this parameter sub-space stability etc. can easily be tested\cite{Peterson2017}.   If the designer had then satisfied themselves that this was a sufficiently interesting design, they could then go forwards with a transfer learning\cite{Humbird2018} or alternative statistical approach\cite{Gopalaswamy2019} to incorporating higher fidelity simulations, and finally experiments.

Finally we note that although the algorithm discussed here is as `blind' as the process that produced the natural world, we speculate that one day it might be possible for the metaheuristic stage in this process to be replaced by a genuine artificial intelligence algorithm that really can imagine and conceive of new designs in the same way that the human brain can.

\begin{acknowledgments}
We thank the referee for helpful comments which have improved the quality of this paper. P.W.H. and S.J.R. acknowledge funding from the Engineering and Physical Sciences Research Council. Many thanks to Warren Garbett at the Atomic Weapons Establishment, Aldermaston, Kris McGlinchey and Jeremy Chittenden at Imperial College, London, and Jim Gaffney, Luc Peterson and Kelli Humbird at Lawrence Livermore National Laboratory for useful discussions. The original idea for this article came from \url{http://boxcar2d.com}.
\end{acknowledgments}

\appendix

\section{Genetic Algorithm Details}

The capsule designs in the starting population are seeded by, for each layer, sampling a material (DT gas, DT ice or CH, each with equal probability), sampling a density to be used if the material is DT gas (uniformly in log-space; the least informative prior for a scale parameter), and sampling a thickness (again uniformly in log-space). If the total radius exceeds the allowed amount the design is scaled down to satisfy the size constraint. The drive designs (as a function of time) in the starting population are expressed as a sum of 15 Gaussians. Each Gaussian has its centre in consecutive 1ns periods of time, and is characterised by three parameters, one for where in the nanosecond the peak occurs, one for the peak temperature, and one for the standard deviation for the Gaussian. Location of peak is sampled uniformly and the component peak temperature and standard deviation are sampled uniformly in log-space. Finally the drive temperature is then uniformly scaled down so that the peak temperature and total energy in the drive are within the prescribed limits, and modified so the drive temperature does not decrease faster than permitted.

The selection method was fitness proportionate selection, with a fitness ($f$) function of $f=Y^\alpha$, with $\alpha=\frac{1}{2}$ chosen. Typically a higher $\alpha$ corresponds to a faster rate of convergence at the cost of being more likely to get caught in local minima. Crossover is performed separately for the capsule and the drive, so that the crossover respects the structure of the problem and is neutral towards drive-capsule interactions. For the capsule an integer $i$ is randomly sampled between the zero and the number of sections (5 here) inclusive, and 1-point crossover is used at that point in the capsule design. Crossover for the drive is done in the same way with the Gaussian components.  Mutations for the capsule are carried out by randomly selecting one of the sections, and with equal probabilities resampling either the material, thickness, or DT gas density in the same was as was used to create the starting population. A similar process is applied to the drive. Finally we use `elitism'; the best in a population automatically survives unmodified to the next generation. 

The heuristic described here is a simple implementation of a genetic algorithm to illustrate the concept that metaheuristics can produce plausible designs from extremely large parameter spaces. It is possible performance can be improved by including more algorithmic features e.g. reducing the mutation rate over time, or by using more complex evolutionary algorithms\cite{Borenstein2014,Hansen2001}.

\nocite{*}
\bibliography{genetic_PoP}% Produces the bibliography via BibTeX.

\end{document}